\documentclass[%
 reprint,
 amsmath,amssymb,
 aps,
]{revtex4-1}

\usepackage{amsthm}
\usepackage{graphicx}
\usepackage{subfigure}
\usepackage{dcolumn}
\usepackage{bm}
\usepackage{float}
\usepackage{booktabs} 
\usepackage{amsmath,amsthm} 
\usepackage{setspace} 
\usepackage[sort&compress]{natbib}
\usepackage{extarrows}

\begin{document}

\title{Generic Quantum Walks with Memory on Regular Graphs}

\author{Dan Li$^{1,2}$}
\email{lidansusu007@163.com}
\author{Michael Mc Gettrick$^{2}$}
\email{michael.mcgettrick@nuigalway.ie}
\author{Fei Gao$^{1}$}
\email{gaof@bupt.edu.cn}
\author{Jie Xu$^{1}$}
\author{Qiao-Yan Wen$^{1}$}
\affiliation{
 $^{1}$State Key Laboratory of Networking and Switching Technology, Beijing University of Posts and Telecommunications, Beijing, 100876, China\\
 $^{2}$The De Brun Centre for Computational Algebra, School of Mathematics, The National University of Ireland, Galway.
}
\date{\today}

\begin{abstract}

Quantum walks with memory(QWM) are a type of modified quantum walks that record the walker's latest path. As we know, only two kinds of QWM are presented up to now.  It is desired to design more QWM for research, so that we can explore  the potential of QWM. In this work, through presenting the one-to-one correspondence between QWM on a regular graph  and quantum walks without memory(QWoM) on line digraph of the regular graph, we construct a generic model of  QWM on regular graphs. This construction gives a general scheme for building all possible standard QWM on regular graphs and makes it possible to study properties of different kinds of QWM. Here, by taking the simplest example which is  QWM with 1 memory on the line, we analyze some properties of QWM, such as variance, occupancy rate and localization.

\begin{description}
\item[PACS numbers]
03.67.Lx, 02.10.Yn, 02.40.Pc
\end{description}

\end{abstract}

\pacs{Valid PACS appear here}

\maketitle

\section{\label{sec:level1}Introduction}

Due to constructive quantum interference along the paths in the discrete or the continuous version, quantum walks provide a method to explore all possible paths in a parallel way. Many kinds of models of quantum walk  have been proposed, such as single-particle quantum walks \cite{119,120,121,122}, two-particle quantum walks \cite{123,124,125}, three-state quantum walks \cite{126,131}, controlled interacting quantum walks  \cite{115,116}, indistinguishable particle quantum walks \cite{127,128}, disordered quantum walk \cite{129,130}, quantum walks on closed surfaces \cite{132}, etc. Each type of quantum walks has its own special features and advantages. Therefore, algorithms based on quantum walks have been established as a dominant technique in quantum computation, ranging from element distinctness \cite{110} to database searching \cite{111,112,113,114}, from constructing quantum Hash schemes \cite{115,116} to graph isomorphism testing \cite{117,118}.

Most quantum walks been studied are quantum walks without memory(QWoM) on regular graphs, such as line, circle and lattice. Quantum walks with memory(QWM) only have been studied in \cite{005,006,007,008}, while classical walks with memory have been used in research on the behavior of hunting, searching and building human memory search model. Standard QWM are a kind of modified quantum walks that have many extra coins to record the walker's latest path. As we know,  there are only two kinds of QWM  presented up to now.  Rohde et al presented a kind of QWM provided by recycled coins and a memory of the coin-flip history \cite{005}. Mc Gettrick presented another kind of QWM whose coin state decides the shift is `Reflect' or `Transmit' \cite{006,007}. Konno and Machida provided limit theorems for Mc Gettrick's QWM\cite{008}. The evolutions of these two QWM accord with our intuition. There doesn't exist a general scheme for building all possible standard QWM. It is desired to design more QWM for research, so that we can explore the potential of QWM. Furthermore, if we want to design a QWM on a complex graph or a position-dependent QWM, it seems impossible to design the unitary evolution for a QWM by intuition.

In this paper, we construct a generic model that includes all possible standard QWM on regular graphs. By analyzing the mathematical formalism of QWoM and two existing QWM, we find that QWM on a regular graph can be transformed into a QWoM on line digraph of the regular graph. Furthermore, the mapping is one-to-one.  That is, we can study QWoM on line digraph of a regular graph instead of the corresponding QWM on the regular graph. There is only one coin for QWoM, while there are at least two coins for  QWM. Therefore, this replacement decreases the coin space  and simplifies the  analytic process for QWM. Then we construct a generic model of  QWoM on  line digraph of regular graphs. This new model is actually the generic model of QWM on regular graphs, and it gives a general scheme for building all possible standard QWM on regular graphs.

With this model, it becomes possible to build any wanted QWM on regular graphs and to study  properties of different kinds of QWM.  In this paper, by taking the simplest example which is QWM with 1 memory on the line, we analyze  some properties of QWM, such as variance, occupancy rate and localization. And we focus on its relation with partition and coin shift function, which are introduced for designing QWM. Through analysis and research, we get some interesting and useful results.

The paper is structured as follows. In Sect.\ref{sec:level2}, we present the one-to-one correspondence between QWM on a regular graph  and QWoM on line digraph of the regular graph. In Sect.\ref{sec:level3}, we construct a generic model that includes all possible standard QWM on regular graphs. Then, in  Sect.\ref{sec:level4}, by taking the simplest example which is QWM with 1 memory on the line, we analyze  some general properties of this kind of QWM. Finally, a short conclusion is given in Sect.\ref{sec:level5}.

\section{\label{sec:level2}Relation between QWM and QWoM}

\theoremstyle{remark}
\newtheorem*{definition}{\indent Definition}
\newtheorem*{observation}{\indent Observation}
\newtheorem{lemma}{\indent Lemma}
\newtheorem*{theorem}{\indent Theorem}
\newtheorem*{corollary}{\indent Corollary}
\newtheorem*{conjecture}{\indent Conjecture}

\def\QEDclosed{\mbox{\rule[0pt]{1.3ex}{1.3ex}}}
\def\QED{\QEDclosed}
\def\proof{\indent{\em Proof}.}
\def\endproof{\hspace*{\fill}~\QED\par\endtrivlist\unskip}

In this part, we introduce the standard formalization of discrete-time QWoM and  two kinds of existing QWM. By analyzing the relation between QWoM and QWM, we show that the evolution of QWM on a regular graph is same with the  evolution of QWoM on  line digraph of the regular graph.

For the standard discrete-time QWoM on a linear graph, the walker is a bipartite system $|x,c\rangle$, where $x$ is the position of the walker in the graph and $c$ is the coin which decide the shift of the walker. The evolution is decomposed into two steps, $U=S*C$, defined as
{\setlength\arraycolsep{2pt}
\begin{eqnarray}\label{Euq001}
     & & C:|x,c\rangle \rightarrow \sum_j A_{c,j} |x,j\rangle,
    \nonumber\\
    &  & S:|x,c\rangle \rightarrow |x+c,c\rangle,
\hspace{1mm}
\end{eqnarray}}where $A$ is a unitary coin matrix defining the transition amplitudes. The coin takes values $\pm 1$ (right or left respectively). After evolving $t$  steps, the output state is $|\psi_{out}\rangle=(SC)^t |\psi_{in}\rangle$.

For QWM, there isn't a generic model which including  all possible standard QWM. Till now, there are only two kinds of QWM as we know.

For QWM in \cite{005}, the evolution is decomposed into two steps, $U=S*C$, defined as
{\setlength\arraycolsep{2pt}
\begin{eqnarray}\label{Euq201}
     & & C:|x,c_1,\cdots,c_{d+1}\rangle \rightarrow \sum_j A_{c_{d+1},j}|x,c_1,\cdots,j\rangle,
    \nonumber\\
    &  & S:|x,c_1,\cdots,c_{d+1}\rangle \rightarrow |x+c_{d+1},c_{d+1},c_1,\cdots,c_d\rangle,
\hspace{1mm}
\end{eqnarray}}where $A$ is still the  unitary coin matrix defining the transition amplitudes. $x$ is the current position and $\{c_i=\pm 1|i=1,\cdots,d\}$ record the shift of the walker $i$ steps before. $c_{d+1}$ is the coin which decides the shift of the walker.

For QWM in \cite{006,007},  the evolution is decomposed into two steps, $U=S*C$, defined as
{\setlength\arraycolsep{2pt}
\begin{eqnarray}\label{Euq202}
     & & C:|x_0,x_1,\cdots,x_d,c\rangle \rightarrow \sum_j A_{c,j} |x_0,x_1,\cdots,x_d,j\rangle,
    \nonumber\\
    &  & S:|x_0,x_1,\cdots,x_d,1\rangle \rightarrow |x_1,x_0,x_1,\cdots,x_{d-1},1\rangle;
    \nonumber\\
    &  &\ \ |x_0,x_1,\cdots,x_d,-1\rangle \rightarrow |2x_0-x_1,x_0,x_1,\cdots,x_{d-1},-1\rangle, \nonumber\\
\hspace{1mm}
\end{eqnarray}}where $A$ is the unitary coin matrix. $x_0$ is the current position and $\{x_i|i=1,\cdots,d\}$ record the positions of the walker $i$ steps before. The coin $c$ takes values $\pm 1$, and $x_{i+1}=x_i\pm 1$.

To build the bridge between QWM and QWoM, we provide a preface for future needs here. We denote by $G=(V,E)$ a digraph with vertex set $V(G)$ and arc set $E(G)$.  With fixed labeling of vertices, the adjacency matrix of a digraph $G$ with N vertices, denoted by $M(G)$, is the $N\times N$ (0,1)-matrix with \textit{ij}-th element defined by $M_{i,j}(G)=1$ if $(x_i,x_j)\in E(G)$ and $M_{i,j}(G)=0$, otherwise. The line digraph of a digraph $G$, denoted by $\overrightarrow{L}G$, is defined as follows: the vertex set of $\overrightarrow{L}G$ is $E(G)$; for $x_a$, $x_b$, $x_c$, $x_d \in V(G)$, $((x_a,x_b),(x_c,x_d))\in E(\overrightarrow{L}G)$ if and only if $(x_a,x_b)$ and $(x_c,x_d)$ are both in $E(G)$ and $x_b=x_c$. The line digraph of $\overrightarrow{L}G$ is denoted by $\overrightarrow{L}^2G$. Similarly, there are $\overrightarrow{L}^dG$s with $d\in N^\ast$. For simplicity, we call all of them line digraph of $G$.

Then we show how to transform  QWM on a regular graph $G$ to QWoM on line digraph of $G$. From the definition of line digraph, we know a vertice of $\overrightarrow{L}^dG$ is a $d$-length path on graph $G$ in fact. Therefore, there is a one-to-one correspondence between $|x,c_1,\cdots,c_d\rangle$ and $|(x-c_1\cdots-c_d, \cdots,x-c_1,x)\rangle$, where $\{(x-c_1\cdots-c_d, \cdots,x-c_1,x)\}$ is the vertex set of $\overrightarrow{L}^dG$. Then, due to the relation between $|x,c_1,\cdots,c_d,c\rangle$ and $|(x-c_1\cdots-c_d, \cdots,x-c_1,x),c\rangle$, we build a bridge between the two kinds of QWM.

For example, we choose the QWM with 2 memory on the line, i.e., 3 qubit coins. For QWM in \cite{005}, the evolution is as follows.
{\setlength\arraycolsep{2pt}
\begin{eqnarray}\label{Euq0051}
    &&|x,c_1,c_2,c\rangle \stackrel{C}{\longrightarrow} \sum_j A_{c,j}|x,c_1,c_2,j\rangle\nonumber\\
     &&\ \ \ \ \ \ \ \ \ \ \ \ \ \ \ \stackrel{S}{\longrightarrow} \sum_j A_{c,j}|x+j,j,c_1,c_2\rangle.
\hspace{1mm}
\end{eqnarray}} It also can be written as
{\setlength\arraycolsep{2pt}
\begin{eqnarray}\label{Euq0052}
    && |(x-c_1-c_2,x-c_1,x),c\rangle  \nonumber\\
    &&\ \ \ \ \ \stackrel{C}{\longrightarrow}\sum_j A_{c,j}|(x-c_1-c_2,x-c_1,x),j\rangle\nonumber\\
    &&\ \ \ \ \ \stackrel{S}{\longrightarrow} \sum_j A_{c,j}|(x-c_1,x,x+j),c_2\rangle.
\hspace{1mm}
\end{eqnarray}}

Eq.\ref{Euq0051} and Eq.\ref{Euq0052} correspond to QWM on $G$ and QWoM on $\overrightarrow{L}^2G$ respectively. We can easily know that the evolutions are essentially the same, because $|x+j,j,c_1,c_2\rangle$ corresponds to $|(x-c_1,x,x+j),c_2\rangle$, where $(x-c_1,x,x+j)$ is a vertex of $\overrightarrow{L}^2G$. In addition, the evolution space $\overrightarrow{L}^2G\otimes H_2$ for QWoM on the line digraph of $G$ is spanned by $\{|(x-c_1-c_2,x-c_1,x),j\rangle|c_1,c_2,j=\pm1\}$, where $H_2$ is the Hilbert space for a $2$-dimensional coin. Therefore, a QWM with 2 memory on $G$ corresponds to a QWoM (which updates the coin state after shift) on the related $\overrightarrow{L}^2G$.  What a more,  a  QWoM (which updates the coin state after shift) on $\overrightarrow{L}^2G$ corresponds to a QWM with 2 memory on $G$.  That means there is a one-to-one correspondence between QWM on $G$ and QWoM on line digraph of $G$. Similarly, there is a one-to-one correspondence between QWM with $d$ memory on any regular graph $G$ and QWoM on the related line digraph $\overrightarrow{L}^dG$.

\section{\label{sec:level3}{QWM on regular graphs}}

From the above part, we know that there is a one-to-one correspondence between QWM on a regular graph $G$ and QWoM on line digraph of $G$. Therefore, the generic model of QWoM on line digraph of $G$, which including all possible QWoM (which updates the coin state after shift) on line digraph of $G$, is actually the generic model of QWM on $G$, which including all possible standard QWM on $G$. Then we only need to construct the generic QWoM on line digraph of $G$ instead of generic QWM on $G$.

Here we introduce two definitions to prepare for constructing the model of generic QWoM on line digraph of an $m$-regular graph $G$ (These two definitions are inspired by \cite{021}). These two definitions are introduced to show the shift of the walker along the graph $G$.

\textbf{Definition 1} Let $G$  an $m$-regular graph. Define $\pi$ be a partition of $\overrightarrow{L}^dG$ such that
\begin{equation}\label{Euq002}
    \pi: \overrightarrow{L}^dG\rightarrow \{ C_1, C_2, \cdots, C_m\},
\end{equation}
where $\{C_k|k=1,\cdots,m\}$ satisfy that $V(C_k)=V(\overrightarrow{L}^dG)$, $\bigcup_kE(C_k)=E(\overrightarrow{L}^dG)$  and for every vertex $v\in V(C_k)$, the outdegree is $1$. Dicycle factorization is a kind of partition which satisfies that for every vertex $v\in V(C_k)$, the outdegree and indegree are 1. We denote the set of partitions of $\overrightarrow{L}^dG$ by $\Pi_{\overrightarrow{L}^dG}$.

We show two partitions in Fig.1. The original graph $G$ is the infinite line in Fig.1(a). The line digraph of $G$, $\overrightarrow{L}G$, is shown in Fig.1(b). We show two partitions $\pi_1$ and $\pi_2$ in Fig.1(b,c) respectively, by using different color lines to denote $C_k$.

\begin{figure}[!ht]\label{Fig001}
 \begin{center}
  \subfigure[]{
\label{Fig001a}
\includegraphics[width=8cm]{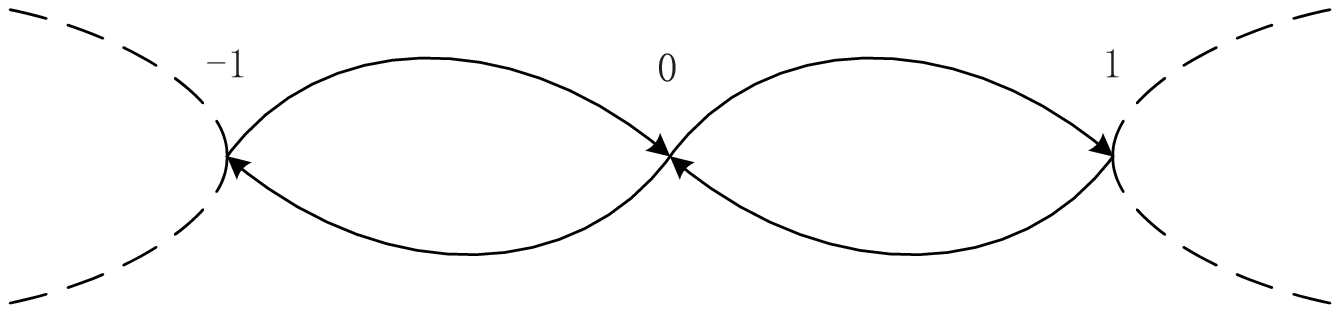}} \\
  \subfigure[]{
  \label{Fig001b}
 \includegraphics[width=8cm]{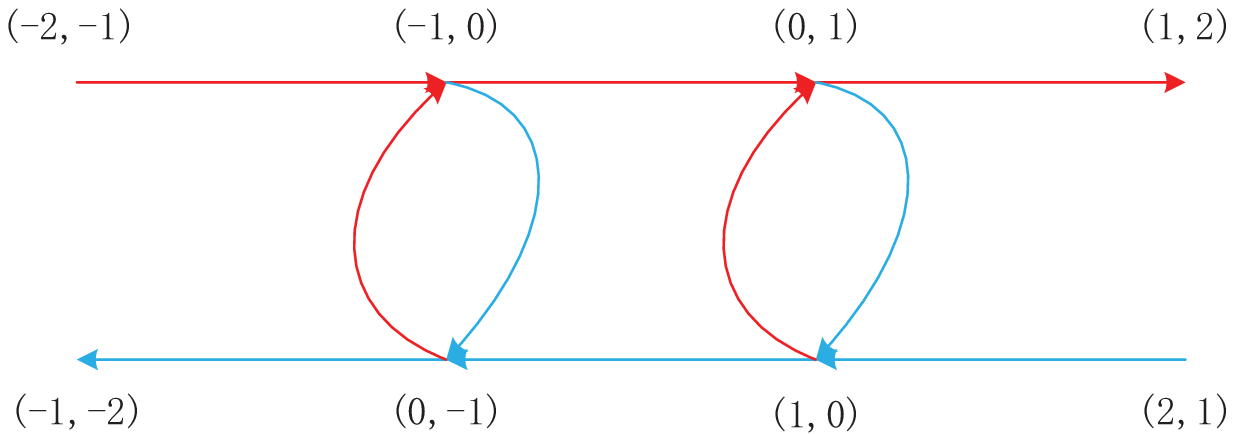}}
  \subfigure[]{
  \label{Fig001c}
\includegraphics[width=8cm]{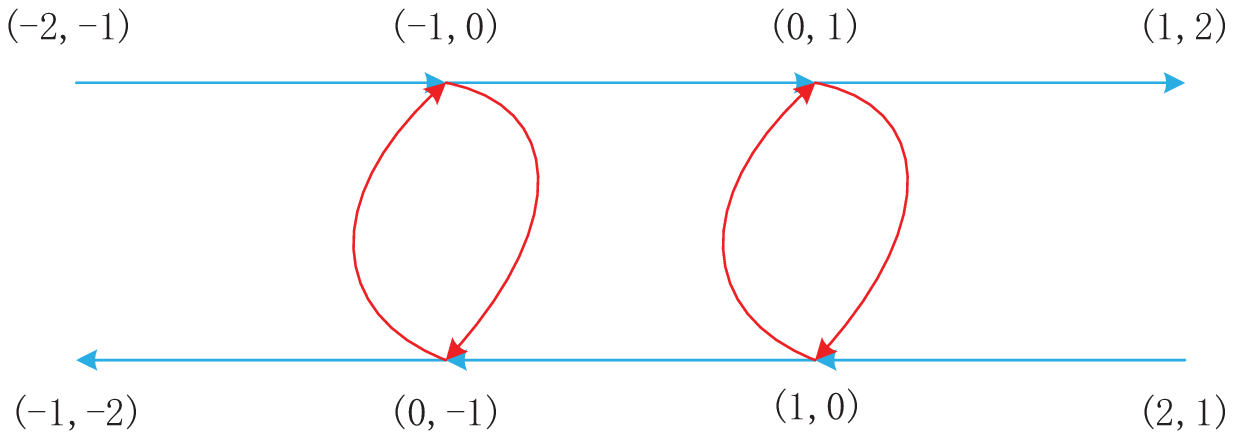}}\\
  \end{center}
\renewcommand{\figurename}{Fig.}
\caption{The original digraph $G$ and the line digraph of $G$ denoted by $\vec{L}G$. Subgraph (b) and (c) show two partitions of $\vec{L}G$ by using different color lines to denote $C_k$. }
\end{figure}

\textbf{Definition 2} For $\pi\in\Pi_{\overrightarrow{L}^dG}$ with $\overrightarrow{L}^dG \stackrel{\pi}{\longrightarrow} \{ C_1, C_2, \cdots, C_m\}$, define
\begin{equation}\label{Euq003}
    f_{C_k}: V(\overrightarrow{L}^dG) \rightarrow V(\overrightarrow{L}^dG)
\end{equation}
such that for any $v\in V(\overrightarrow{L}^dG)$,
\begin{equation}\label{Euq004}
    (v,f_{C_k}(v))\in E(C_k)
\end{equation}

In what follows, we construct the generic QWoM on $\overrightarrow{L}^{d}G$, i.e., generic model of QWM with $d$-step memory on an $m$-regular graph $G$.

\textbf{Definition 3} For a QWoM  on the line digraph of $G$ denoted by $\overrightarrow{L}^{d}G$, the evolution is decomposed into two steps, $U=S*C$, defined as
{\setlength\arraycolsep{2pt}
\begin{eqnarray} \label{Euq006}
     C: &&|v,c\rangle \longrightarrow \sum_j A_{c,c_j} |v,c_j\rangle;\nonumber\\
     S: &&|v,c_j\rangle \longrightarrow |f_{C_j}(v),gc(v,c_j)\rangle
\hspace{1mm}
\end{eqnarray}}where $A$ is a unitary coin matrix defining the transition amplitudes, which may be position-time-dependent. $v$ is the position at $\overrightarrow{L}G$. The coin $c$ decides the shift of the walker. The coin shift function $gc$ is defined as follows:
\begin{equation}\label{Euq007}
    gc: V(\overrightarrow{L}^{d}G)\otimes \textit{H}_m \longrightarrow \textit{H}_m,
\end{equation}where $\textit{H}_m$ is the Hilbert space for an m-dimensional coin, spanned by $\{c_1,c_2,\cdots,c_m\}$.

The shift operator $S$ dictates the walker to walk along the subgraph $C_k$ when the coin is $c_k$. The coin shift function $gc$ updates the coin after moving. We should remind readers that QWM is decided by coin operator, partition and coin shift function.

Till now, we get a generic model of QWoM on line digraph of $G$, i.e., a generic model of QWM on regular graph $G$. However, the model seems too formalized. Next, we will show the concrete form of the model. It is worth reminding that the evolution of quantum walks on an infinite graph at time $t$ equals the evolution of quantum walks on a bigger finite graph at time $t$. Therefore, we only need to consider finite graph when we focus on the outstate after $t$ steps.

The coin operator $C$ for this new model is similar with that for QWoM, which may be position-history-dependent. However, the shift operator $S$ for this model is more complex. Below we will show the concrete form of the shift operator.

Suppose $M(G)$ is the adjacent matrix of an m-regular graph $G$ on $N$ vertices. Then a partition of $G$ is actually a partition of the adjacent matrix $M(G)$, i.e.,
\begin{equation}\label{Euq093}
   M(G)\xlongequal{\quad\pi\quad} M(D_1)+M(D_2)+\cdots+M(D_m),
\end{equation}
where there is exactly one entry which is 1 and 0 elsewhere for every row vector of $M(D_k)$, $k=1,\cdots,m$. Furthermore, a dicycle factorization is a special partition that there is exactly one entry which is 1 and 0 elsewhere for every column vector of $M(D_k)$, $k=1,\cdots,m$.

Then, we introduce an important theorem \cite{091} about the adjacent matrix of the line digraph of $G$.

\textbf{Theorem 1}\label{T1} Let $G$ be an $m$-regular digraph on $N$ vertices and let $\{D_1,D_2,\cdots,D_m\}$ be a dicycle factorization of $G$. Then there is a labeling of $V(\overrightarrow{L}G)$ such that
\begin{equation}\label{Euq091}
    M(\overrightarrow{L}G)=\left(
  \begin{array}{cccc}
    M(D_1) & M(D_2) & \cdots & M(D_m) \\
    M(D_1) & M(D_2) & \cdots & M(D_m) \\
    \vdots & \vdots & \ddots & \vdots \\
    M(D_1) & M(D_2) & \cdots & M(D_m) \\
  \end{array}
\right).
\end{equation}

With this theorem, we can get the adjacent matrix $M(\overrightarrow{L}^dG)$ for any $m$-regular graph $G$. At first, we need to partition the $m$-regular graph $G$ into $\{D_1,D_2,\cdots,D_m\}$ which is a dicycle factorization of $G$. This process is very easy for a regular graph. According to Theorem 1, we can get  the adjacent matrix $M(\overrightarrow{L}G)$. Then we  partition $M(\overrightarrow{L}G)$ to
{\setlength\arraycolsep{2pt}
\begin{eqnarray} \label{Euq092}
      M(\overrightarrow{L}G)&&=\left(
  \begin{array}{cccc}
    M(D_1) &  & \ &  \\
     & M(D_2) & & \\
     &  & \ddots &  \\
     &  &  & M(D_m) \\
  \end{array}
\right)\nonumber\\
&&+\cdots+\left(
  \begin{array}{cccc}
     0& M(D_2) &  &\\
    & 0&M(D_3) & \\
     &  & 0& M(D_m) \\
     M(D_1) &  & & 0 \\
  \end{array}
\right),\nonumber\\
\hspace{1mm}
\end{eqnarray}}where the formalization of each term in the sum is similar to that of determinant. This partition is still a dicycle factorization of $M(\overrightarrow{L}G)$. We can get the adjacent matrix $M(\overrightarrow{L}^2G)$ by using Theorem 1 again. Iterating this process, we can get the adjacent matrixes of $\overrightarrow{L}^dG$ for any $d$. The vertices of  $V(\overrightarrow{L}^dG)$ are labeled by  $v_1,\cdots, v_{N\cdot 2^{d-1}},\cdots v_{N\cdot 2^{d}}$.

Till now, we can get the adjacent matrix of $\overrightarrow{L}^dG$ for any $d$ and any regular graph $G$. A partition of $\overrightarrow{L}^dG$ is actually a partition of the adjacent matrix of $\overrightarrow{L}^dG$, i.e.,
\begin{equation}\label{Euq093}
   M(\overrightarrow{L}^dG)\xlongequal{\quad\pi\quad} M(C_1)+M(C_2)+\cdots+M(C_m),
\end{equation}
where there is exactly one entry which is 1 and 0 elsewhere for every row vector of $M(C_k)$, $k=1,\cdots,m$. Furthermore, $M(C_k)_{(v_i,v_j)}=1$ if and only if $f_{C_k}(v_i)=v_j$. Therefore, the shift operator $S$ is
\begin{equation}\label{Euq094}
   S=\sum_{i,k} M(C_k)^T |v_i\rangle\langle v_i| \otimes |gc(v_i,c_k)\rangle\langle c_k|.
\end{equation}

From the above equation, we find that in order to ensure the unitarity of shift operator $S$, $gc$ has to follow some rules.  From Theorem 1, we can easily know that $M(\overrightarrow{L}^dG)$ can be viewed as a combination of $m$ same block matrixes, i.e.,
\begin{equation}\label{Euq0916}
   M(\overrightarrow{L}^dG)=\left(
                              \begin{array}{ccc}
                                M_1(\overrightarrow{L}^dG)  \\
                                \vdots  \\
                                M_m(\overrightarrow{L}^dG)  \\
                              \end{array}
                            \right)
\end{equation}where each of $\{M_i|i=1,\cdots,m\}$ includes $N\cdot m^{d-1}$ rows of $M(\overrightarrow{L}^dG)$. Furthermore, because we choose  dicycle factorizations to partition the regular graph $G$ and its line digraphs, there is exactly one entry which is 1 and 0 elsewhere for every column vector of $M_i(\overrightarrow{L}^dG)$. Therefore, there are $m$ $\{C_k,v_i\}$s satisfy $v=f_{C_k}(v_i)$ for any $v\in V(\overrightarrow{L}^dG)$. Under the sort order of vertices, in order to make sure the unitarity of the shift operator $S$, for the corresponding $m$ $\{c_k,v_i\}$s, the set of $\{gc(v_i,c_k)\}$ has to satisfy
{\setlength\arraycolsep{2pt}
\begin{eqnarray} \label{Euq908}
\{\overbrace{gc(v_i,c_k),\cdots}^m\}=\{c_1,c_2,\cdots,c_m\}.
\hspace{1mm}
\end{eqnarray}}

So far, we finally get the concrete form of generic model of QWoM on line digraph of $G$. Due to the one-to-one correspondence between QWoM on line digraph of $G$ and QWM on $G$, this model is actually the generic model of QWM, which includes all possible standard QWM on regular graphs. More importantly, this model provides a way to build any wanted QWM directly. Furthermore, it seems the walker has to walk on a more complex graph, but the generic model does not increase nor decrease the evolution space. It provides  a fresh viewpoint to study the QWM by transforming the coin space to position space.


\section{\label{sec:level4}QWM with 1 memory on the line}

Generic model of QWM provides a way to design any wanted QWM. With all standard QWM on regular graphs, it is possible to study  properties of different kinds of QWM. However, it is unrealistic to study each particular situation in this paper. In this part, we focus on properties of the most simple one, QWM with 1 memory on the line, i.e., $m=2,d=1$, and its relation with partition and coin shift function.

In order to study in depth  properties of the generic model of QWM, we consider QWM with different partitions and coin shift functions, and the standard QWoM for comparison. We denote the partitions and coin shift functions in \cite{005,006} $\pi_1$,$\pi_2$,$gc_1$,$gc_2$ respectively. We also consider a random partition $\pi_3$ and a random dicycle factorization partition $\pi_4$. Therefore, there are 6 kinds of QWM: QWM with $\pi_1$,$gc_1$; QWM with $\pi_2$,$gc_1$; QWM with $\pi_2$,$gc_2$; QWM with $\pi_3$,$gc_1$; QWM with $\pi_4$,$gc_1$; QWM with $\pi_4$,$gc_2$. To avoid the confusion that QWM we simulated include QWM with $\pi_2, gc_1$, but not QWM with $\pi_1, gc_2$, we remind readers that for QWM with partition $\pi_1$, the coin shift function $gc_2$ does not satisfy the constraint Eq.\ref{Euq908}. We leave details for Appendix A. Based on the same reason, there doesn't exist QWM with $\pi_3$,$gc_2$. For the coin operator, in this paper, we only consider the Hadamard matrix as the coin operator.

QWM bring possibilities for new phenomena. However, QWM with a dicycle factorization partition and the coin shift function $gc_2$ reduces the possibilities. When the initial position state is $|0\rangle_p$, there are QWM with different dicycle factorization partitions and the coin shift function $gc_2$ which create same probability distribution for any initial coin state. The evolution of this kind of QWM is only affected by partition at three positions around the center, i.e. -1, 0, 1. Therefore, the number of different QWM with a dicycle factorization partition and coin shift function $gc_2$ is $8=2^3$. Luckily, this phenomenon does not appear in other kinds of QWM. For other kinds of QWM, the number of different QWM increases exponentially with time $t$, as we predicted.

\begin{figure}[!ht]\label{Fig002}
  \begin{center}
  \includegraphics[width=8cm]{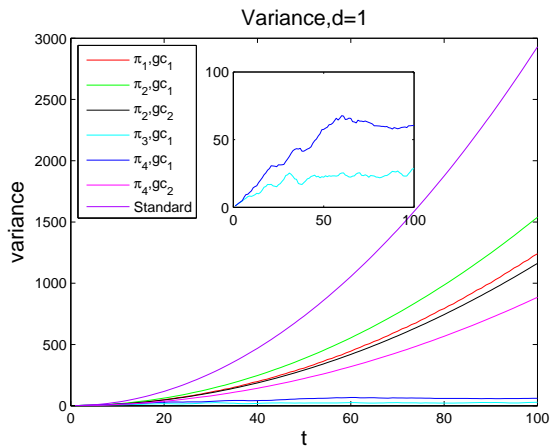}\\
   \end{center}
  \renewcommand{\figurename}{Fig.}
  \caption{Variance of  QWM. The inset shows the variance of QWM with $\pi_3,gc_1$ and  $\pi_4,gc_1$ once again.  }
\end{figure}

Now we consider variance of QWM. For QWoM, we usually only consider current position of the walker, and ignore the coin state. For QWM, we still only consider current position rather than the memory. Therefore,  variance is defined as follows,
{\setlength\arraycolsep{2pt}
\begin{eqnarray} \label{Euq203}
   variance=\sum_xp(x)\cdot x^2-\left[\sum_xp(x)\cdot x\right]^2,\nonumber
\hspace{1mm}
\end{eqnarray}}where $x$ means current position of the walker, $p(x)$ means the norm's square of amplitude for the walker at position $x$. Ellinas and Smyrnakis \cite{019} have shown that the general form for variance of a quantum walk is $\sigma(t)^2=K_2t^2+K_1t+K_0^2$. QWoM still follow the rule. From Fig.2, we know that most QWM are ballistic, except for QWM with a random partition  and the coin shift function $gc_1$. It is surprising that  QWM with a random dicycle factorization partition and $gc_2$ are still ballistic while the other QWM with a random partition are diffusive. We find that random partition destroys the ballistic nature of QWM; only QWM with a dicycle factorization and the coin function $gc_2$ remain ballistic. From another angle, QWM with the coin shift function $gc_2$ can generate ballistic behavior. For QWM with the coin shift function $gc_1$, only organized partitions (such as partition with repetition) can help to make QWM ballistic. Luckily, considering application of QWM, only QWM with a organized partition will  gain extensive attention of researchers.

\begin{figure}[!ht]\label{Fig003}
 \begin{center}
  \includegraphics[width=8cm]{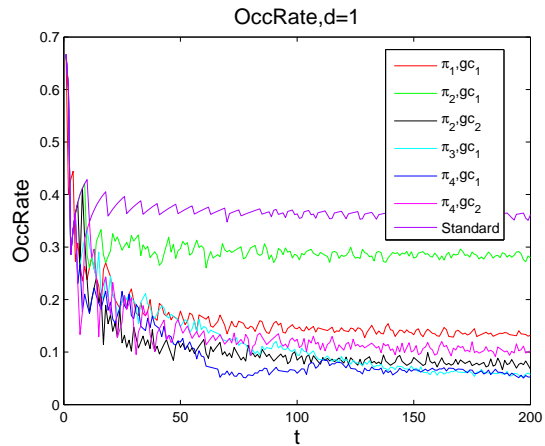}\\
  \end{center}
  \renewcommand{\figurename}{Fig.}
  \caption{Occupance rate of QWM.}
\end{figure}

We consider  occupancy rate of QWM here. QWM with a random partition and $gc_1$ are diffusive, but that doesn't  mean there does not exist any quantum property for this kind of QWM. The Occupancy Rate was proposed in \cite{004} as a way of measuring  statistical property of probability distribution. A non-zero value of the occupancy rate for a quantum walk on an infinite graph can be seen as a sign of quantum feature. If the walker has range $N$,  the occupancy rate is defined as
\begin{equation}\label{Euq107}
    OccRate(N,t)=\frac{\#\{x|P(x,t)\geq\frac{1}{N}\}}{N}.
\end{equation} For QWoM on the line \cite{004}, variance has the order $O(t^2)$, and occupancy rate has the order $O(1)$. In addition, for classical walks on the line, variance has the order $O(t)$, and occupancy rate converges to 0.  For QWM in Fig. 3, the order of the occupancy rate is still $O(1)$ for all cases, while the order of variance of QWM with a random partition and the coin shift function $gc_1$ is $O(t)$. This means that QWM do not lose all of their quantum properties even with a random partition.

Localization is an important feature of QWM(by localization, we mean the existence of position $x$ where the asymptotic probability value is non-zero). For QWM, localization is a more common property than for other kinds of quantum walks. Almost every kind of QWM we examine, except QWM with the partition $\pi_2$ and the coin shift function $gc_1$, possesses localization. For QWM with a random partition and the coin shift function $gc_1$, the probability at the origin vibrates sharply because of randomness of the partition. Nevertheless, this kind of QWM still have a high probability at the origin for large values of the time $t$. It shows again the quantum property of QWM with a random partition.

During the research, we find a rare and interesting result. Even though there exist QWM with different partitions which could produce same probability distribution, they  still belong to QWM. Mostly, different kinds of quantum walks produce different probability distributions. There is only one special case. In \cite{009}, Franco et.al find the nonlocalized case of the spatial density probability of the two-dimensional Grover walk can be obtained using only a two-dimensional coin space and a quantum walk in alternate directions. Here, we find another example. The symmetric probability distribution of the standard QWoM on the line with initial state $|0\rangle(\frac{1}{\sqrt{2}}|1\rangle+\frac{i}{\sqrt{2}}|-1\rangle$ can be obtained by the QWM with $\pi_2$ and $gc_2$ when the initial state is $\frac{1}{2}|-1,0\rangle|1\rangle-\frac{1}{2}|-1,0\rangle|-1\rangle-\frac{1}{2}|1,0\rangle|1\rangle+\frac{1}{2}|1,0\rangle|-1\rangle$.
Let us denote the coefficients in the decomposition of the standard QWoM and that of the QWM as $\alpha_{x,c}$ and $\beta_{x,x\pm1,c}$ respectively. Then we have a correspondence between them given by
{\setlength\arraycolsep{2pt}
\begin{eqnarray} \label{Euq108}
     \alpha_{x,1}^{t}&=& (-1)^{\frac{t+x}{2}} e^{i\frac{\pi}{4}} (-i\beta_{x-1,x,1}^t-\beta_{x-1,x,-1}^t); \nonumber\\
       \alpha_{x,-1}^{t}&=& (-1)^{\frac{t+x}{2}} e^{i\frac{\pi}{4}} (-\beta_{x+1,x,1}^t+i\beta_{x+1,x,-1}^t).
\hspace{1mm}
\end{eqnarray}}We leave the proof for Appendix B. Therefore, these two totally different quantum walks produce same probability distribution, i.e.,
{\setlength\arraycolsep{2pt}
\begin{eqnarray} \label{Euq109}
P_x^t&&=|\alpha_{x,1}^t|^2+|\alpha_{x,-1}^{t}|^2 \nonumber\\
&&=|\beta_{x-1,x,1}^t|^2+|\beta_{x-1,x,-1}^t|^2  \nonumber\\
&&\ \ \ +|\beta_{x+1,x,1}^t|^2+|\beta_{x+1,x,-1}^t|^2.
\hspace{1mm}
\end{eqnarray}}

\section{\label{sec:level5}summary}

In this paper, we find the one-to-one correspondence  between QWM with $d$ memory on a regular graph $G$ and QWoM on the associated line digraph  $\overrightarrow{L}^dG$. Through this correspondence, we can study QWoM on the line digraph of $G$ instead of QWM on  $G$. Furthermore, through this correspondence, we construct a generic model which includes all possible standard QWM on regular graphs.

The generic model  may not make the calculation of QWM simpler, because it transforms the coin space to position space, which does not increase or decrease the resource to execute QWM. However, the new model gives a fresh viewpoint to study QWM on regular graphs. Furthermore, this model  gives a general scheme for building all possible standard QWM on regular graphs. Then we can design any required QWM on regular graphs.

What is more, with the generic model of QWM on regular graphs, it is possible to study  properties of different kinds of QWM. Because it is unrealistic to study each particular situation in this paper, we focus on the simplest case which is QWM with 1 memory on the line. We pay most attention to QWM with different kinds of partition and coin shift functions. In this paper, we get the following results:

1. QWM with a sorted partition  have ballistic evolution, while QWM with a random partition may become diffusive. At the same time, QWM with a random dicycle factorization partition and $gc_2$ (which number 8 when we fix the initial position state to $|0\rangle$) are still ballistic. With this result, we can build a ballistic QWM as wish by choosing appropriate partition and coin shift function. We also know that we don't need to study all QWM with a dicycle factorization partition and the coin shift function $gc_2$, because they can be reduced to 8 QWM. Therefore, research of that 8 QWM is enough.

2. QWM have nonzero value of the occupancy rate, even for QWM with a random partition. This means QWM still have quantum property even with a random partition.

3. Localization is a common feature for QWM, but it does not necessarily occur for all QWM.  QWM  with the partition $\pi_2$ and the coin shift function $gc_1$ don't  possess localization property. Our results tell us  which kind of partition and coin shift function we should choose if we want to build a QWM with or without localization.

4. A QWM could produce the same probability distribution as that of a standard QWoM on the line when the initial state is $\sqrt{\frac{1}{2}}|0\rangle_p(|1\rangle_c+i|-1\rangle_c)$. This result may be not useful,  but interesting considering how rare it is.

Our work extends current research on quantum walks by exhibiting a generic model of QWM. Furthermore, the generic model opens a door to the research of QWM by giving a general scheme for constructing all possible standard QWM. We anticipate that the abundant phenomena of QWM will be useful in quantum computation and quantum simulation.

\section*{\label{sec:level6}Appendix}

\section*{\label{sec:level61}A}

First, we show the partitions of the two kinds of QWM \cite{005,006}, labeled by $\pi_1$ and $\pi_2$, in Fig.\ref{Fig001b} and Fig.\ref{Fig001c} respectively. The essential difference between the two partitions is that the partition $\pi_2$ is a  dicycle factorization of  the line digraph in Fig.\ref{Fig001a}, which means that for every vertex $v\in V(C_k)$ $(k=1,2)$,  the outdegree and indegree of $v$ are both 1. This difference makes the coin shift function $gc$ for different QWM in different forms.

We denote the vertices as $v_1,\cdots,v_N,\cdots v_{2N}$(the sort of order should obey the theorem 1 at Sect.\ref{sec:level3}). According to equation \ref{Euq0916}, for any $v\in\overrightarrow{L}G$,  $M(\overrightarrow{L}G)_{(v_i,v)}=1$ if and only if $M(\overrightarrow{L}G)_{(v_{i+N},v)}=1$.

If the partition is a dicycle factorization, because the indegree of any $v$ is 1, there doesn't exist $(v_j,C_k)$ satisfies that $f_{C_k}(v_j)=f_{C_k}(v_{j+N})$.  Therefore, the constraint \ref{Euq908} can be written as
{\setlength\arraycolsep{2pt}
\begin{eqnarray} \label{Euq095}
     gc(v_i,1)+gc(v_{i+N},-1)&&=0; \nonumber\\
      gc(v_i,-1)+gc(v_{i+N},1)&&=0.
\hspace{1mm}
\end{eqnarray}}where $i\leq N\cdot 2^{d-1}$.

If the partition is not a dicycle factorization, there exist $(v_j,C_k)$ satisfies $f_{C}(v_j)=f_{C}(v_{j+N})$. Therefore,
{\setlength\arraycolsep{2pt}
\begin{eqnarray} \label{Euq096}
     gc(v_i,1)+gc(v_{i+N},-1)&&=0; (i\neq j)\nonumber\\
      gc(v_i,-1)+gc(v_{i+N},1)&&=0; (i\neq j)\nonumber\\
      gc(v_j,1)+gc(v_{j+N},1)&&=0;\nonumber\\
       gc(v_j,-1)+gc(v_{j+N},-1)&&=0.
\hspace{1mm}
\end{eqnarray}}Summarize above conditions, the choice
{\setlength\arraycolsep{2pt}
\begin{eqnarray} \label{Euq097}
   gc(v_i,1)=k_i, gc(v_{i+N},1)=-k_i, \nonumber\\
     gc(v_i,-1)=k_i, gc(v_{i+N},-1)=-k_i.
\hspace{1mm}
\end{eqnarray}}with $k_i=\pm1$ befits any partition.

For QWM in \cite{005}, for any $v_j$, there exists $C_k$ such that $f_{C_k}(v_j)=f_{C_k}(v_{j+N})$ and the coin state takes values $\pm 1$. The coin shift function $gc_1$ in \cite{005} satisfies Equ.\ref{Euq097}, which befits any partition.  For QWM in \cite{006}, because $\pi_2$ is a dicycle factorization, there does not exist $(v_j, C_k)$ such that $f_{C_k}(v_j)=f_{C_k}(v_{j+N})$. The coin shift function $gc_2$ in \cite{006} is
{\setlength\arraycolsep{2pt}
\begin{eqnarray} \label{Euq010}
   &&gc(v_i,1)=k_i, \ \ \ \ \ \ gc(v_{i+N},1)=k_i, \nonumber\\
   &&gc(v_i,-1)=-k_i, \ gc(v_{i+N},-1)=-k_i,
\hspace{1mm}
\end{eqnarray}}where $i\leq N$. The $gc_2$ only satisfies the Eqs.\ref{Euq095} rather than Eqs.\ref{Euq096}, which means $gc_2$ only work for QWM with dicycle factorization partition.

\section*{\label{sec:level62}B}

For QWM with $\pi_2$ and $gc_2$,
{\setlength\arraycolsep{2pt}
\begin{eqnarray} \label{Euq098}
     \beta_{x,x+1,1}^{t+1}= \frac{1}{\sqrt{2}}(\beta_{x+1,x,1}^t+\beta_{x+1,x,-1}^t);  \nonumber\\
     \beta_{x,x-1,1}^{t+1}= \frac{1}{\sqrt{2}}(\beta_{x-1,x,1}^t+\beta_{x-1,x,-1}^t); \nonumber\\
     \beta_{x,x-1,-1}^{t+1}=\frac{1}{\sqrt{2}}(\beta_{x+1,x,1}^t-\beta_{x+1,x,-1}^t); \nonumber\\
     \beta_{x,x+1,-1}^{t+1}= \frac{1}{\sqrt{2}}(\beta_{x-1,x,1}^t-\beta_{x-1,x,-1}^t).
\hspace{1mm}
\end{eqnarray}} We first want to prove the amplitudes satisfy the constraint
{\setlength\arraycolsep{2pt}
\begin{eqnarray} \label{Euq099}
     \beta_{x,x+1,1}^t+\beta_{x,x+1,-1}^t+\beta_{x,x-1,1}^t+\beta_{x,x-1,-1}^t&&=0; \nonumber\\
     \beta_{x+1,x,1}^t+\beta_{x-1,x,1}^t&&=0.
\hspace{1mm}
\end{eqnarray}}Our proof works by induction on t. When the QWM begin with the initial state $\frac{1}{2}|-1,0\rangle|1\rangle-\frac{1}{2}|-1,0\rangle|-1\rangle-\frac{1}{2}|1,0\rangle|1\rangle+\frac{1}{2}|1,0\rangle|-1\rangle$,  i.e.,
{\setlength\arraycolsep{2pt}
\begin{eqnarray} \label{Euq0910}
    &&\beta_{-1,0,1}^0 =\frac{1}{2}; \ \ \ \ \ \ \  \beta_{-1,0,-1}^0=-\frac{1}{2};\nonumber\\
    &&\beta_{1,0,1}^0=-\frac{1}{2}; \ \ \ \ \ \ \   \beta_{1,0,-1}^0=\frac{1}{2},
\hspace{1mm}
\end{eqnarray}}It is easy to verify, by means of a direct calculation, that Eqs.\ref{Euq099} is satisfied at $t=0$. Now, we assume the  Eqs.\ref{Euq099} is true for any $x$ at  time t, then we prove that it holds at time $t+1$.
{\setlength\arraycolsep{2pt}
\begin{eqnarray} \label{Euq0911}
    &&\beta_{x,x+1,1}^{t+1}+\beta_{x,x+1,-1}^{t+1}+\beta_{x,x-1,1}^{t+1}+\beta_{x,x-1,-1}^{t+1}\nonumber\\
    &&\ \ =\sqrt{2}(\beta_{x+1,x,1}^t+\beta_{x-1,x,1}^t)\nonumber\\
    &&\ \ =\beta_{x,x+1,1}^{t-1}+\beta_{x,x+1,-1}^{t-1}+\beta_{x,x-1,1}^{t-1}+\beta_{x,x-1,-1}^{t-1}\nonumber\\
    &&\ \ =\sqrt{2}(\beta_{x+1,x,1}^{t-2}+\beta_{x-1,x,1}^{t-2}).
\hspace{1mm}
\end{eqnarray}} Therefore, Eqs.\ref{Euq099} is satisfied at any time $t$.

For the standard QWoM on the line,
{\setlength\arraycolsep{2pt}
\begin{eqnarray} \label{Euq0912}
     \alpha_{x,1}^{t+1}&=& \frac{1}{\sqrt{2}}(\alpha_{x-1,1}^t+\alpha_{x-1,-1}^t); \nonumber\\
       \alpha_{x,-1}^{t+1}&=& \frac{1}{\sqrt{2}}(\alpha_{x+1,1}^t-\alpha_{x+1,-1}^t).
\hspace{1mm}
\end{eqnarray}}We want to prove the relation
{\setlength\arraycolsep{2pt}
\begin{eqnarray} \label{Euq0913}
     \alpha_{x,1}^{t}&=& (-1)^{\frac{t+x}{2}} e^{i\frac{\pi}{4}} (-i\beta_{x-1,x,1}^t-\beta_{x-1,x,-1}^t); \nonumber\\
       \alpha_{x,-1}^{t}&=& (-1)^{\frac{t+x}{2}} e^{i\frac{\pi}{4}} (-\beta_{x+1,x,1}^t+i\beta_{x+1,x,-1}^t),
\hspace{1mm}
\end{eqnarray}}with the initial condition for the standard QWoM given by $\alpha_{0,1}=\frac{1}{\sqrt{2}}$, $\alpha_{0,-1}=\frac{i}{\sqrt{2}}$. Again, we proceed by induction in $t$. When $t=0$, Eqs.\ref{Euq0913} is satisfied by means of a direct calculation. Now, we assume the  Eqs.\ref{Euq0913} is true for any $x$ at  time t, then we prove that it holds at time $t+1$.
{\setlength\arraycolsep{2pt}
\begin{eqnarray} \label{Euq0914}
\alpha_{x,1}^{t+1}&&= \frac{1}{\sqrt{2}}(\alpha_{x-1,1}^t+\alpha_{x-1,-1}^t) \nonumber\\
     &&=\frac{1}{\sqrt{2}}(-1)^{\frac{t+x-1}{2}} e^{i\frac{\pi}{4}} (-i\beta_{x-2,x-1,1}^t-\beta_{x-2,x-1,-1}^t\nonumber\\
     &&\ \ \ \ \ \ -\beta_{x,x-1,1}^t+i\beta_{x,x-1,-1}^t)\nonumber\\
     &&=\frac{1}{\sqrt{2}} (-1)^{\frac{t+x-1}{2}} e^{i\frac{\pi}{4}} (i\beta_{x,x-1,1}^t-\beta_{x-2,x-1,-1}^t\nonumber\\
     &&\ \ \ \ \ \ +\beta_{x-2,x-1,1}^t+i\beta_{x,x-1,-1}^t)\nonumber\\
     &&=(-1)^{\frac{t+x+1}{2}} e^{i\frac{\pi}{4}}(-i\beta_{x-1,x,1}^{t+1}-\beta_{x-1,x,-1}^{t+1});
\end{eqnarray}}
{\setlength\arraycolsep{2pt}
\begin{eqnarray} \label{Euq0915}
\alpha_{x,-1}^{t+1}&&= \frac{1}{\sqrt{2}}(\alpha_{x+1,1}^t-\alpha_{x+1,-1}^t) \nonumber\\
     &&=\frac{1}{\sqrt{2}}(-1)^{\frac{t+x+1}{2}} e^{i\frac{\pi}{4}} (-i\beta_{x,x+1,1}^t-\beta_{x,x+1,-1}^t\nonumber\\
     &&\ \ \ \ \ \ +\beta_{x+2,x+1,1}^t-i\beta_{x+2,x+1,-1}^t)\nonumber\\
     &&=\frac{1}{\sqrt{2}} (-1)^{\frac{t+x+1}{2}} e^{i\frac{\pi}{4}} (i\beta_{x+2,x+1,1}^t-\beta_{x,x+1,-1}^t\nonumber\\
     &&\ \ \ \ \ \ -\beta_{x,x+1,1}^t-i\beta_{x+2,x+1,-1}^t)\nonumber\\
     &&=(-1)^{\frac{t+x+1}{2}} e^{i\frac{\pi}{4}}(-\beta_{x+1,x,1}^{t+1}+i\beta_{x+1,x,-1}^{t+1}).
\end{eqnarray}}Therefore, Eqs. \ref{Euq0913} is satisfied at any time $t$.

\begin{acknowledgments}
This work is supported by NSFC (Grant Nos. 61272057, 61572081), Beijing Higher Education Young Elite Teacher Project (Grant Nos. YETP0475, YETP0477), BUPT Excellent Ph.D. Students Foundation(Grant Nos. CX201326), China Scholarship Council(Grant Nos. 201306470046).
\end{acknowledgments}

\end{document}